\documentclass[11pt,twoside]{article}
\usepackage{asp2004}
\usepackage{psfig}
\usepackage{epsf}
\usepackage{graphics}
\usepackage{graphicx}
\usepackage{lscape}
\def\edcomment#1{\iffalse\marginpar{\raggedright\sl#1\/}\else\relax\fi}
\reversemarginpar
\begin{document}
\title{From Clark Lake to Chandra:\\
Closing in on the Low End of the
Relativistic Electron Spectra in Extragalactic Sources}
\author{D. E. Harris}
\affil{Harvard-Smithsonian Center for Astrophysics, 60 Garden St.,
  Cambridge, MA 02138}

\begin{abstract}
The limited angular resolutions and sensitivities historically
available below 300 MHz have made it difficult to define the low end
of the electron energy distribution, N($\gamma$).  We extrapolate down
from the well observed segments of radio spectra with almost complete
ignorance of what N($\gamma$) is actually doing.  We do not know if
there is a low energy cutoff or if there are other deviations from
extrapolated power laws.  The result is that we really do not have a
good estimate of the total energy density and pressure of the
relativistic plasmas we study.  The situation is even worse for
Inverse Compton (IC) X-ray emission, several flavors of which rely on
electrons of Lorentz factors, $\gamma$, of 1000, 300, or in some cases
of order 50.  If our assumed extrapolations are wrong, some IC
emission models may have to be abandoned.  We present several examples
and demonstrate that the Long Wave Array (LWA) should have sufficient
sensitivity and resolution to obtain meaningful constraints on
N($\gamma$) at low energies.
\end{abstract}
\thispagestyle{plain}


\section{Introduction}

At this meeting, we are honoring a man who has not lost sight of his
life-long passion for low frequency radio astronomy.  Erickson's
exploits are legendary and described in detail in other contributions
in this volume.  We note that twenty years ago, November 1984,
Erickson and Cane organized a Green Bank Workshop on Low Frequency
Radio Astronomy and that 10 of the 43 participants of the '84 workshop
managed to survive the intervening 20 years and are registered at this
meeting.  In section~\ref{sec:cl} we review what sort of science we
were doing at that time, from the point of view of a guest observer
with the TPT at the Clark Lake Radio Observatory.

In addition to celebrating Erickson's birthday, we have come to
evaluate and nourish the LWA proposal.  In my opinion, if the LWA does
nothing more than tell us something new about the low end of
relativistic electron spectra, it will have been worth the effort!
Thus, in section~\ref{sec:ne} we review some of the reasons why these
low energy electrons are important.  Briefly, most of the energy
density resides below $\gamma\approx$5000, i.e. the low end of the
electron energy distribution which is generally not sampled by radio
observations with arcsec synthesized beams.

In section~\ref{sec:ic}, we focus on several of the IC
processes proposed for X-ray emission which rely on the presence of
electrons with $\gamma\leq$~1000.  What is normally not discussed in
the literature is the extrapolation required from the 'radio-sampled'
section of N($\gamma$) to the energies required for the particular IC
model being proposed.  We consider various examples of IC X-ray
sources and show what LWA resolutions and frequency coverage could do
to mitigate the uncertainty of the current extrapolations.

We use the standard definition of the spectral index, $\alpha$: flux
density,\\
S$_{\nu}\propto\nu^{-\alpha}$.

\section{What were we doing at Clark Lake anyway?\label{sec:cl}}

The aspect of our previous work at Clark Lake which most relates to my
topic here was published in Harris et al. (1988), a multi-wavelength
study of 4 fields containing steep spectrum radio sources.  We used
the Clark Lake TPT at 30 and 57 MHz, the VLA at 1.4 and 5 GHz, the
Westerbork Synthesis Radio Telescope at 610 MHz, the Dominion Radio
Astrophysical Observatory at 1.4 GHz, and the Einstein
Observatory for X-ray data.

\begin{figure}[!ht]
\begin{center}
\plottwo{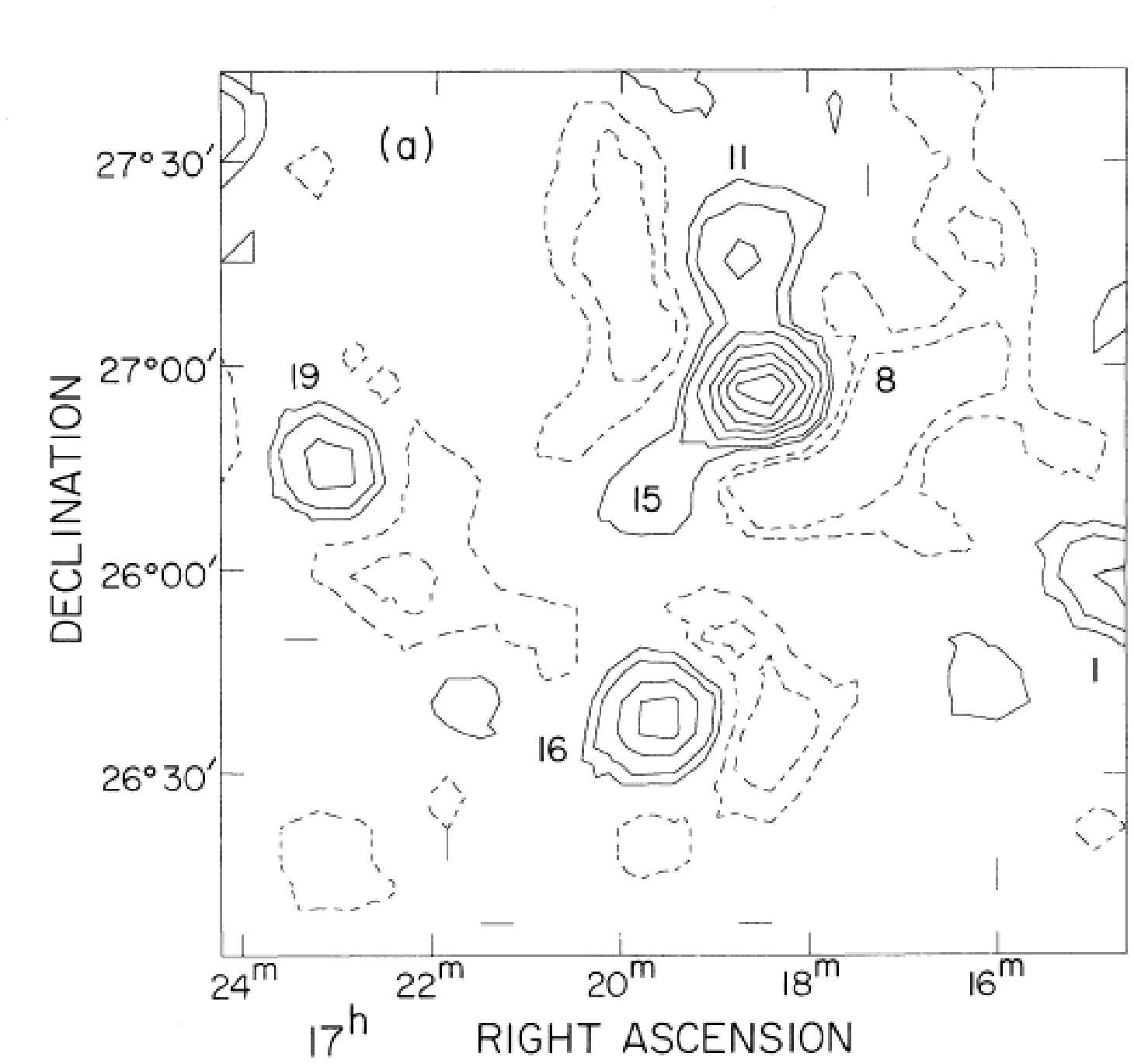}{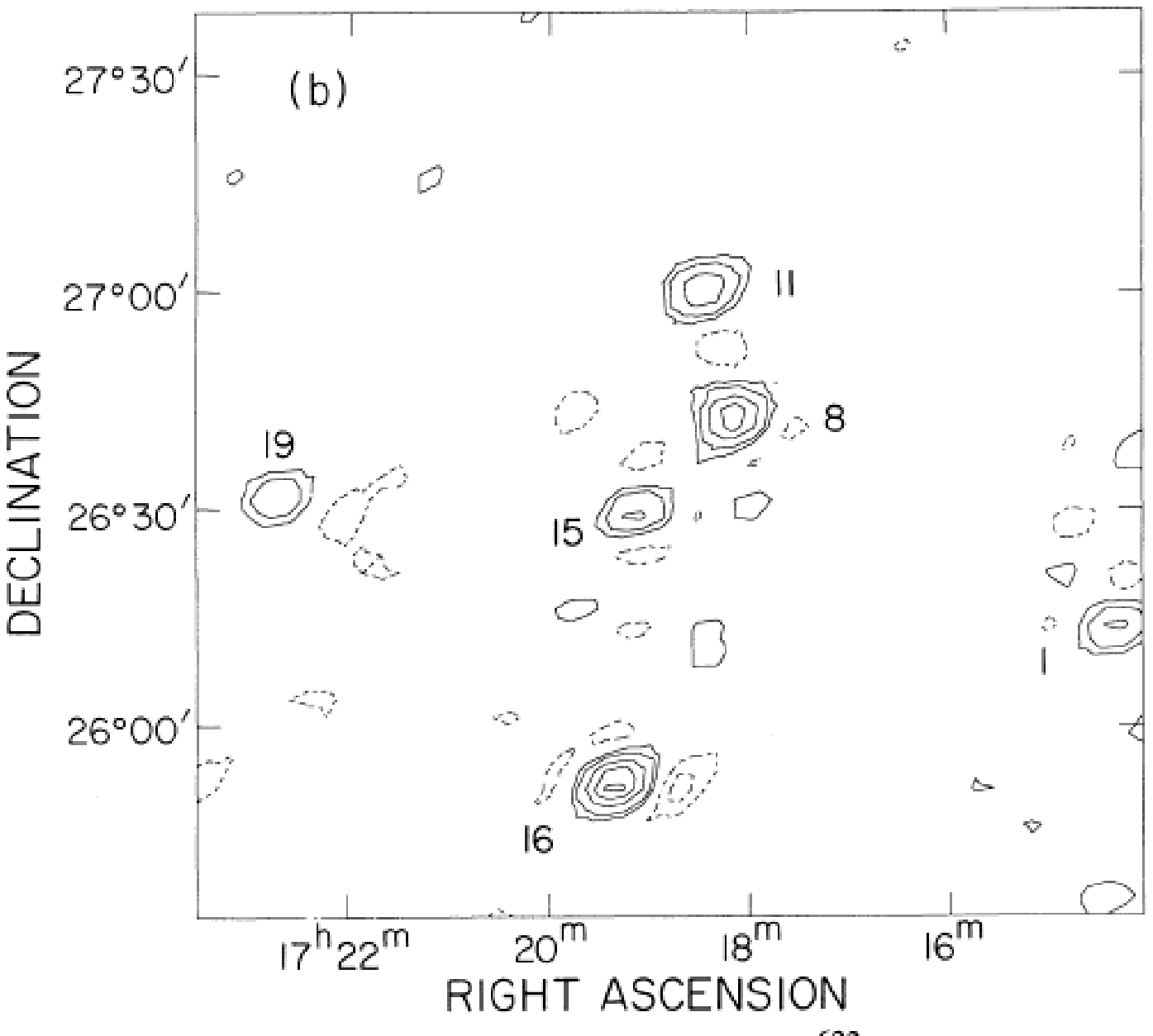}

\caption{A copy of contour diagrams at 30 (left) and 57 MHz (right)
  from the TPT at Clark Lake.  Contour levels are $\pm1, \pm$2, 4, 8,
  10, and 12 Jy/beam.  Note the dominating intensity of source 8 at 30
  MHz.  These figures originally appeared in Harris et
  al. (1988)\label{fig:clcon}}
\end{center}
\end{figure}

The beauty of Clark Lake data was that it made it easy to locate the
steep spectrum source since the high flux density at 26 MHz used to
select these fields was determined with a beam of 14$'\times92'$.  As
seen in fig.~\ref{fig:clcon}, the source of interest is clearly
indicated by the 30MHz map.  The 57 Mhz positions were good enough to
tie into the higher frequency data so that we could then determine an
optical identification, define the source morphology, and obtain radio
spectra from 30 to 5000 MHz.

Once we had the source distance, size, and spectrum, we could run the usual
numbers for synchrotron models (Pacholczyk, 1970), and obtain equipartition
field strengths, relevant energy densities, etc.

A result from that paper was the extrapolation of the electron spectra.
We showed that if you believe the extrapolation, the number of
electrons in these sources between $\gamma$=100 to 1000 is comparable
to that in Cygnus A.  If you extend the spectrum down to $\gamma$=10,
then there are more low energy electrons in these relatively weak FRI
sources than in Cygnus A. These low energy electrons serve as a
measure of the integrated luminosity of the source over its active
life since they have a very long lifetime.

\section{General considerations concerning the low end of N($\gamma$)\label{sec:ne}}

\subsection{Uncertainties in the low end of N($\gamma$)}
 
What happens to N($\gamma$) below $\gamma$=a few thousand where we can
begin to estimate it from ground based radio observations?
Possibilities include:

\begin{itemize}

\item{$\alpha$ increases to values between 1 and 2 (or more) so that
  there are many more low energy electrons than we would predict by
  extrapolating the observed spectrum.  For examples of sources with
  low frequency steep spectra, see Braude et al.  (1969) and Roger et
  al. (1973).}

\item{$\alpha$ remains essentially constant as observed in the cm band
so that calculated extrapolations down to $\gamma\approx$10 give the
proper number of low energy electrons.}

\item{The spectrum flattens towards $\alpha$=0 so that there are
actually fewer low energy electrons than predicted by the
extrapolation.}
	
\item{there is a low energy cutoff between $\gamma$=300 and 1800 so
that there are essentially no electrons at very low energy.}

\end{itemize}

In spite of our almost total ignorance as to which of these conditions
occurs, many IC models of X-ray emission rely on the assumption that
an extrapolation to low $\gamma$'s based on the observed $\alpha_r$,
is valid.  The uncertainty as to the form of N($\gamma$) also
compromises our attempts to determine if the plasma is pair dominated
or consists of protons and electrons.  Using pressure balance
arguments between non-thermal pressures in radio lobes and the ambient
medium in clusters of galaxies, we often find that the external
thermal pressure is larger than the minimum non thermal pressure.
This could indicate that the relativistic plasma is far from equipartition
between field and particles; that the filling factor is much less than
1; that the low energy electrons we are discussing have not been
properly estimated; and/or that there is a significant contribution to
the energy density from relativistic protons (the ``1+k term'').
Minimizing our uncertainty on the contribution to the pressure from
the low energy electrons serves to strengthen constraints on the
proton contribution to the non-thermal pressure.

\subsection{Dependence of synchrotron parameters on the lower
integration limit of the frequency}

\begin{figure}[!ht] 
\plottwo{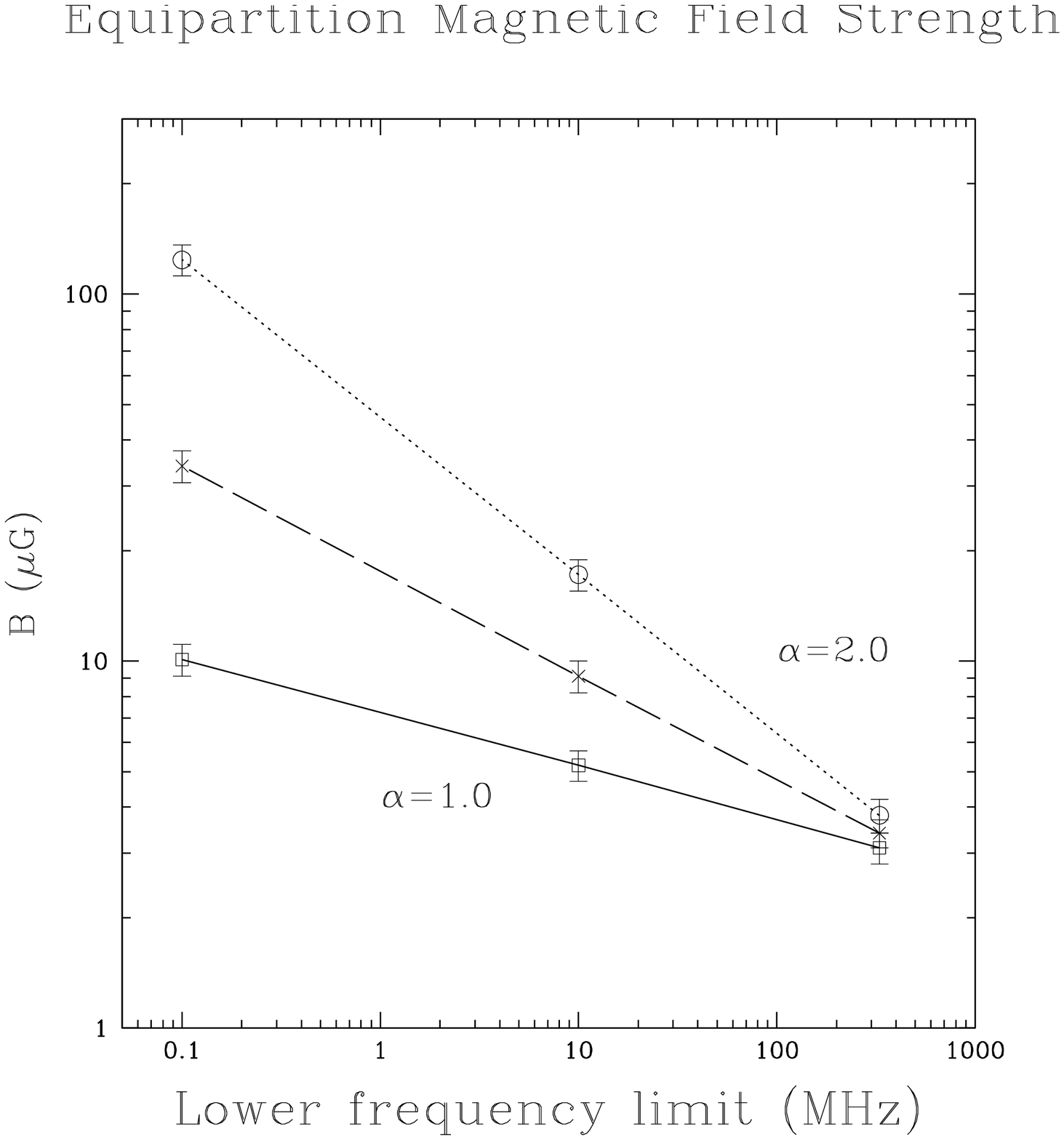}{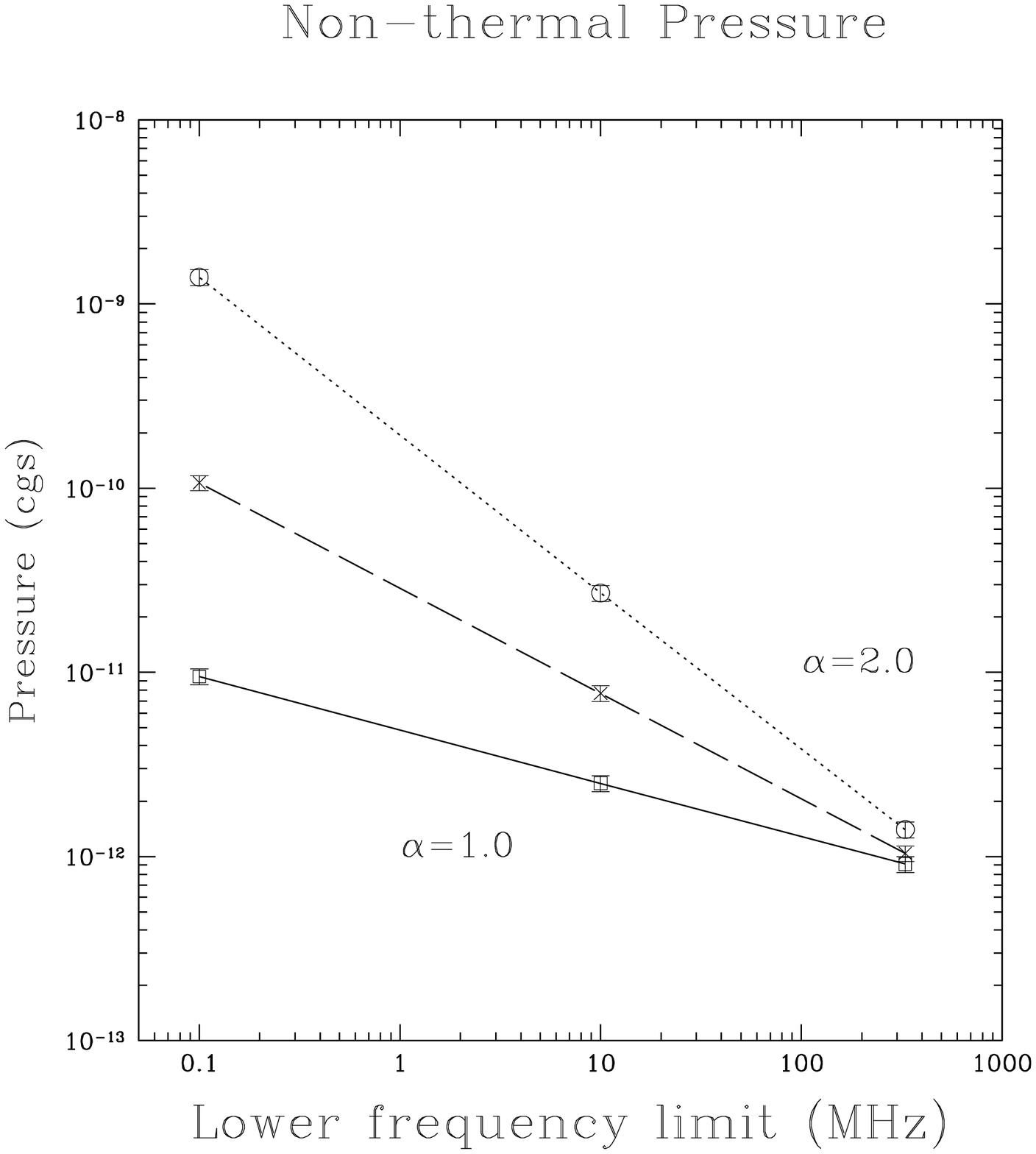}

\caption{The dependence of the equipartition field strength and the
  non-thermal pressure on the lower limit of the radio spectrum.  The
  three lines correspond to values of $\alpha$ of 1.0 (solid), 1.5
  (dashed), and 2.0 (dotted).  The actual flux density used
  corresponds to that from the north lobe of 3C 351.  The left panel
  is the equipartition magnetic field strength and the right panel is
  the non-thermal pressure.\label{fig:lowcut1}}

\end{figure}

\begin{figure}[!ht] 
\plottwo{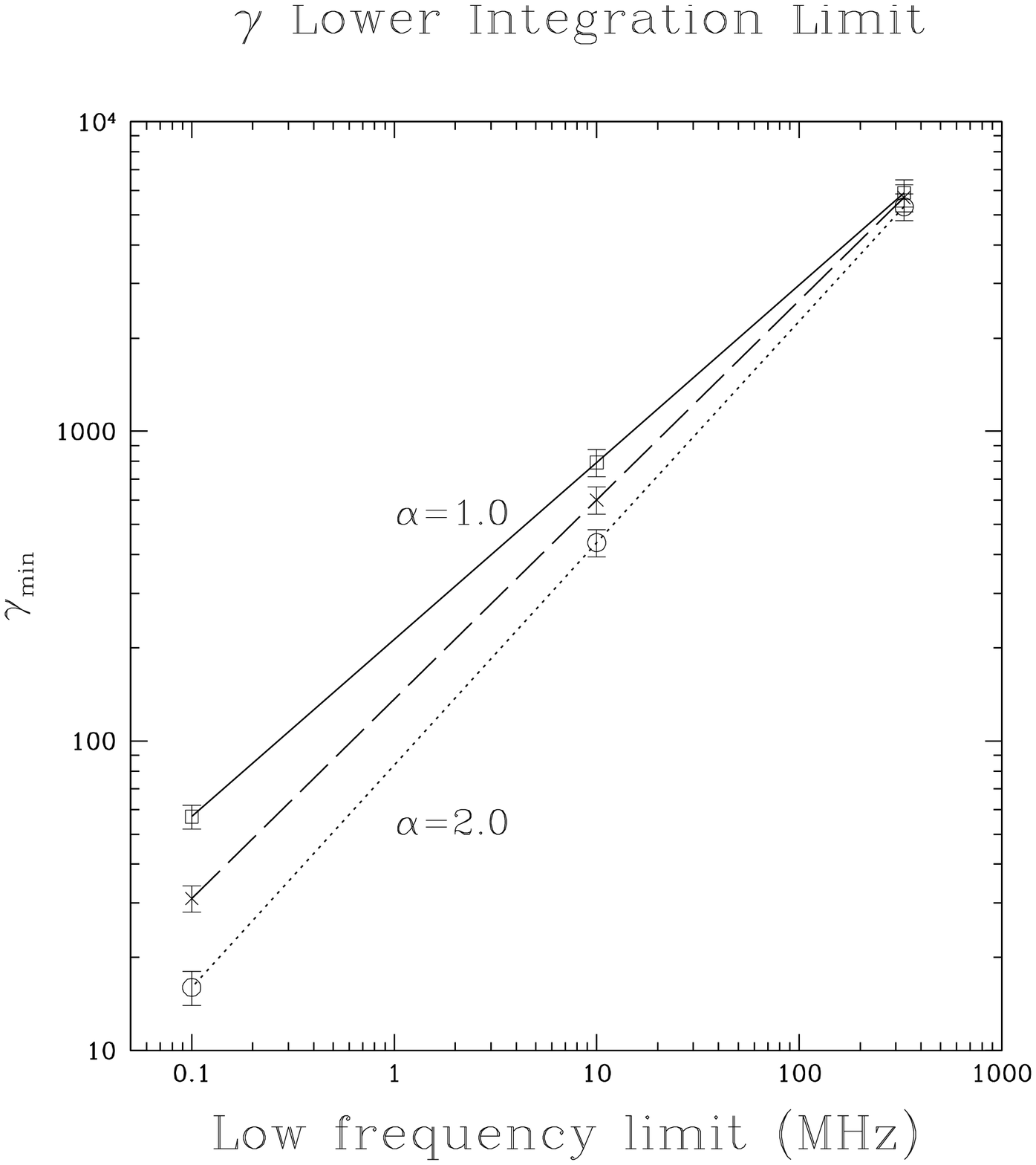}{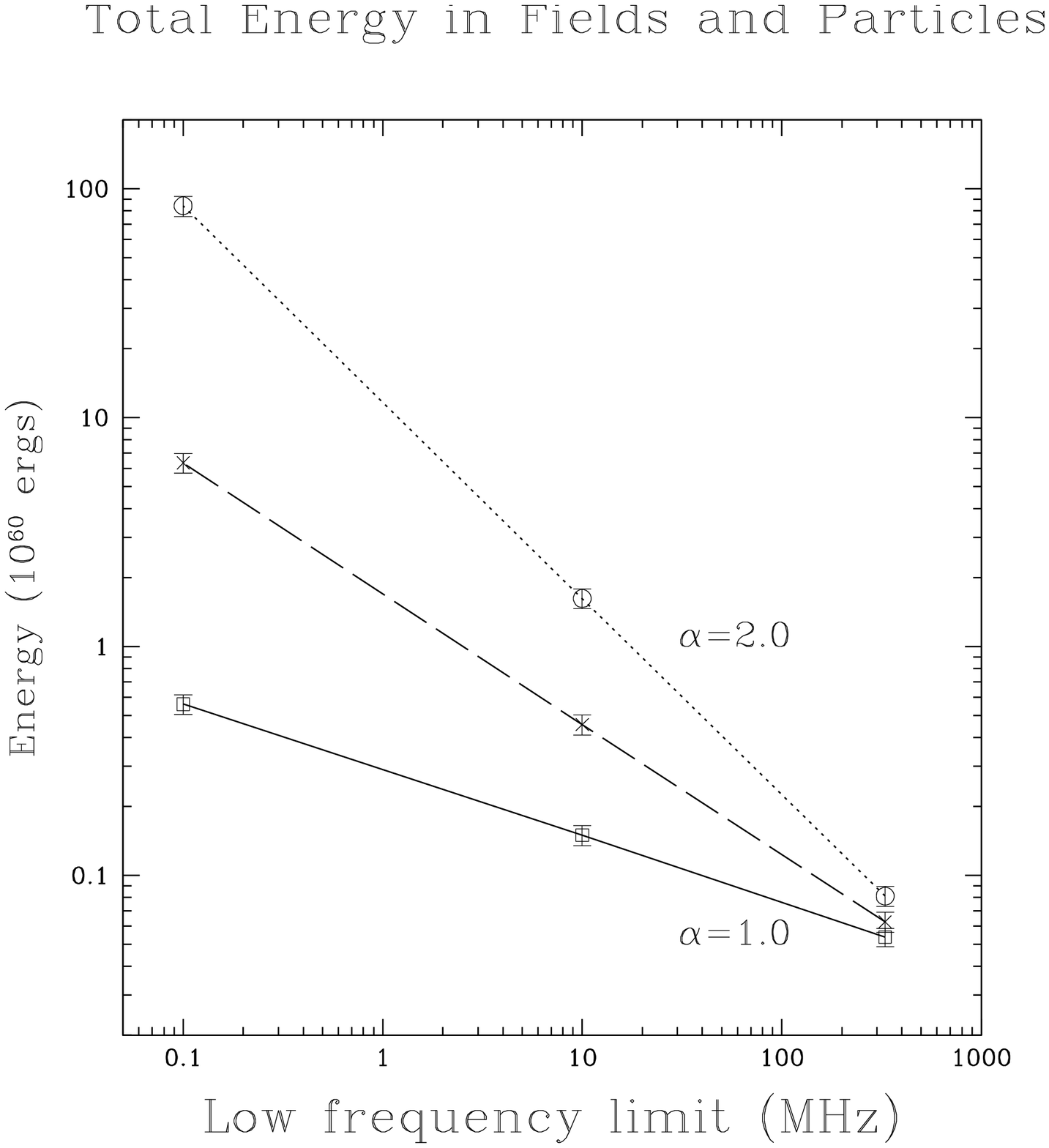}

\caption{The dependence of $\gamma$ and total energy on the lower
integration frequency.  The left panel shows the value of $\gamma$ of
the electrons radiating at the given radio frequency.  The right panel
shows the total energy stored in particles and
fields.\label{fig:lowcut2}}

\end{figure}

To illustrate what we are dealing with, consider the differences
resulting from choosing a low frequency limit of integration of 330,
10, and 0.1 MHz.  In the following examples, we use the measured flux
density at 1.4 GHz of the north lobe of 3C351 with a filling factor,
$\phi$=1, assumed equipartition, and k=0 (no significant contribution
from relativistic protons).  In figs.~\ref{fig:lowcut1} and
~\ref{fig:lowcut2}, as a function of the low frequency limit of
integration, we show the values of $\gamma$; the equipartition field
strength, B$_{eq}$; the minimum non-thermal pressure; and the total
energy in particles and fields for values of $\alpha$ of 1, 1.5, and
2.

Naturally, B$_{eq}$ goes up to keep pace with the increasing particle
energy density as the lower integration frequency decreases and the
pressure follows B$_{eq}$.

\subsection{What is the evidence for a turnover or cutoff in N($\gamma$)?}

In the case of shock acceleration for plasmas consisting of electrons
and protons, there has been a long standing problem of how to get
electrons up to the $\gamma$'s required for diffusive shock acceleration
to be effective.  Lesch and Birk (1997) argue that this 'injection
energy' is 1800, although other analyses produce lower values
(e.g. 100 to 800, Eilek \& Hughes 1991). If these conditions
(i.e. protons are present in equal numbers with electrons) pertain to
the generation of synchrotron emitting plasmas, there would be little
expectation that the observed power law of the radio spectrum would be
normally related to the behavior of N($\gamma$) for $\gamma~<$~1800.

Observationally, Carilli et al. (1991) when fitting synchrotron models
to hotspots of Cygnus A had to invoke a cutoff in N($\gamma$) at about
$\gamma$=400 in order to accommodate the measured flux densities at 151
and 327 MHz (B$_{eq}$=300 $\mu$G).  SSA explanations suffer from an
outrageously large B, or very small filling factor.

Both of these arguments lend credence to the notion that we should not
expect our assumed extrapolations to small values of $\gamma$ to be valid.

\section{The IC connection to synchrotron emitting plasmas\label{sec:ic}}

Here we focus on IC emission in the X-ray band of 0.2-10 keV, but
there are of course many considerations in the literature of IC
emission in other bands (e.g. to explain cluster excesses in the
extreme ultraviolet and to explain hard (20 to 80 keV) X-rays
from clusters of galaxies.

Synchrotron emission depends on the energy density of the magnetic
field, u(B) and IC emission depends on the energy density of the
photons, u($\nu$).  Both of these are essentially mandatory processes,
but IC is 'more' mandatory in the sense that every radio source must
have at least a photon energy density from the cosmic microwave
background, u(CMB) and the observed u(sync), whereas there is no
a priori knowledge of the average value of the magnetic field strength
or of its spatial variations.

In this section we consider several flavors of IC emission and give a
few examples of particular sources and demonstrate how data from the
LWA would provide critical sampling of the underlying electron spectra
at low energies.  For illustrative purposes, we take the LWA
resolution to be 15$"$, 5$"$, and 2$"$ (FWHM of synthesized beam) at
10, 30, and 75 MHz, respectively.  The straw man sensitivity at these 3
frequencies is assumed to be 3, 1.6, and 1 mJy (point source
sensitivity for 1 hour integration, 4 MHz bandwidth, and 1
polarization).

\subsection{Synchrotron Self-Compton (SSC) Emission}

This is the well known process used in blazar cores and in the
brightest radio hotspots (Harris, Carilli, and Perley, 1994;
Hardcastle et al. 2004) to model X-ray and higher frequency emissions.
Unless field strengths are much greater than the equipartition fields
estimated, the electrons responsible for the observed X-rays have
energies comparable to those required for the observed synchrotron
spectrum (typical $\gamma$'s of 5000 and greater), so they are not of
direct interest here (no extrapolation of the electron spectrum is
involved in deriving model parameters).

\subsection{IC/CMB Emission from Radio Lobes}

Achieving robust detections of IC/CMB X-rays with good signal to noise
from radio lobes has been fairly elusive observationally even though
we know the emission has to exist.  There are however, a growing
number of lobe detections with Chandra which generally are in
agreement with equipartition field estimates to within a factor of a
few.  For the soft X-ray band, electron energies of
$\gamma~\approx$~1000 are required, regardless of the redshift of the
source (the shift in the peak of the CMB spectrum compensates for the
shift in the observed frequency).

\begin{figure}[!ht] 
\plottwo{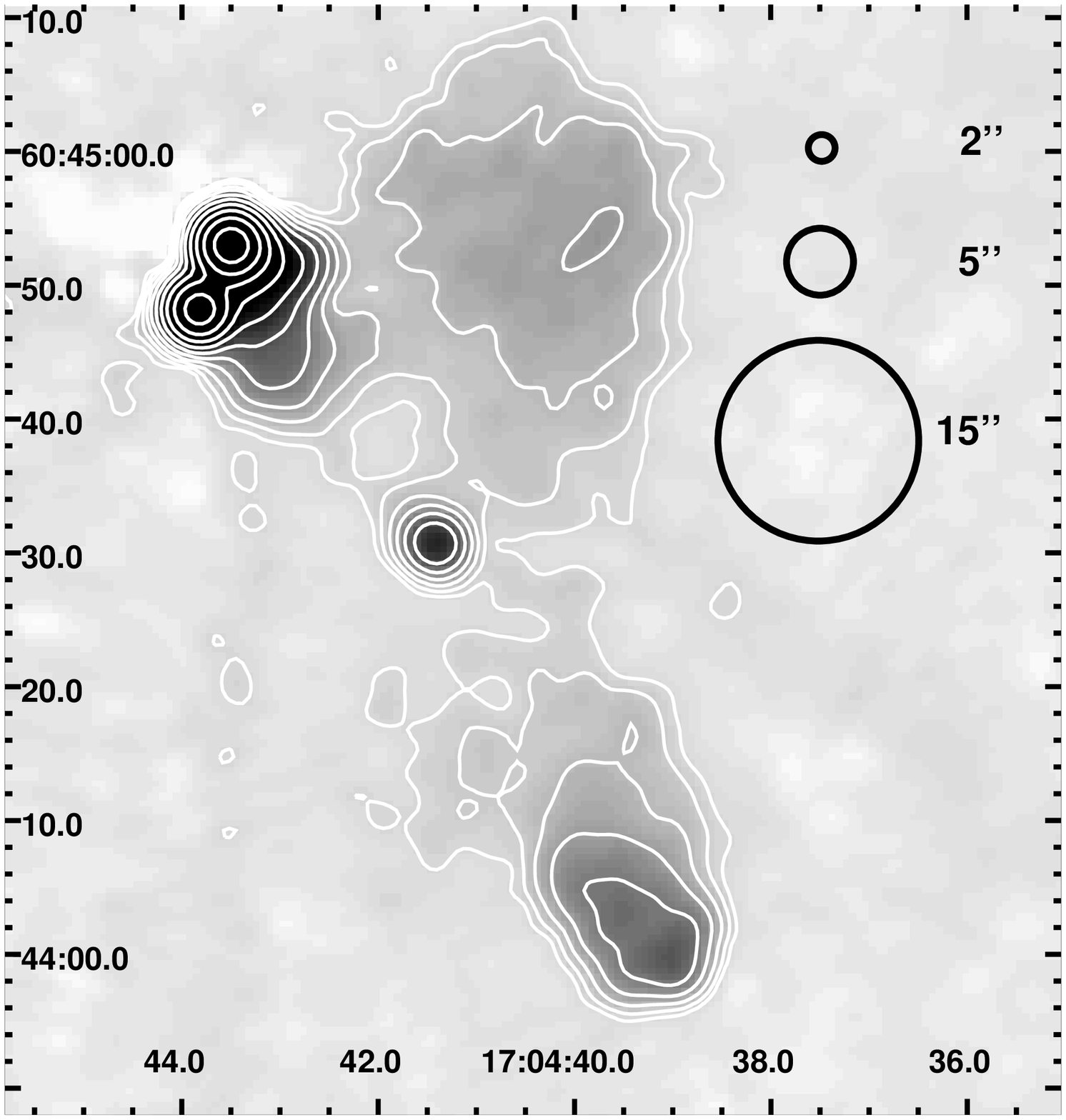}{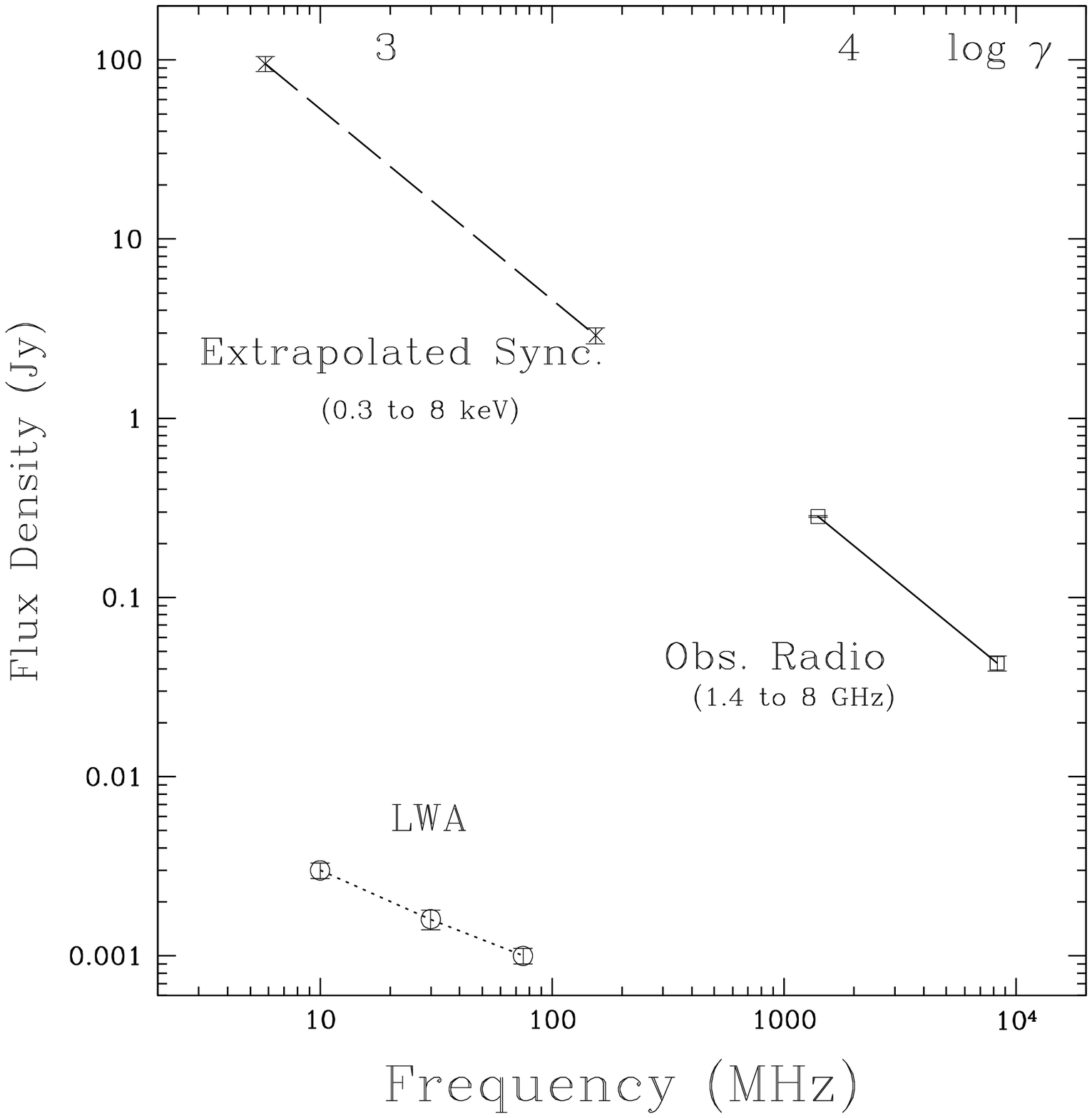}
\caption{3C351, a quasar at z=0.371.  The left panel shows an 8GHz
image from the VLA with FWHM beamsize of 3$"$ (Hardcastle et
al. 2002).  Also shown are the 3 FWHM beamsizes of the LWA at 75, 30,
and 10 MHz (2$"$, 5$"$, and 15$"$, respectively).  Contour levels
start at 0.2 mJy/beam and increase by factors of two.  The right panel
shows the observed segment of the radio spectrum, the extrapolated
segment ($\alpha_r$=1.06) which would produce the IC/CMB X-rays
between 0.3 and 8 keV, and the point source sensitivity expected for
the LWA.  Along the top edge are shown the values of log$\gamma$
corresponding to the equipartition field strength of
6$\mu$G.\label{fig:351}}

\end{figure}

The first example is the northern lobe of 3C 351 shown in
fig.~\ref{fig:351}, for which Hardcastle et al. (2002) report a
detection consisting of 59$\pm$14 net counts (0.4-7 keV).  Since some
kind soul has seen to it that the very bright (both in radio and
X-rays) hotspots have been positioned well away from the lobe, even
the lowest frequency beam of the LWA will be sufficient to measure the
lobe intensity.

From the spectrum, it is clear that the LWA sensitivity should provide
good s/n even if the spectrum flattens at low frequencies.  It is also
obvious that the LWA radio data will sample most of the segment of the
electron spectrum which is supposed to generate the IC/CMB X-ray
emission.  Thus the LWA will have sufficient resolution and
sensitivity to provide a critical test of the IC/CMB model.  The usual
train of arguments will then be tightened to give better values of the
average magnetic field strength (i.e. the uncertainty as to the
validity of the extrapolation of N($\gamma$) will no longer be
present).

\begin{figure}[!ht]
\plottwo{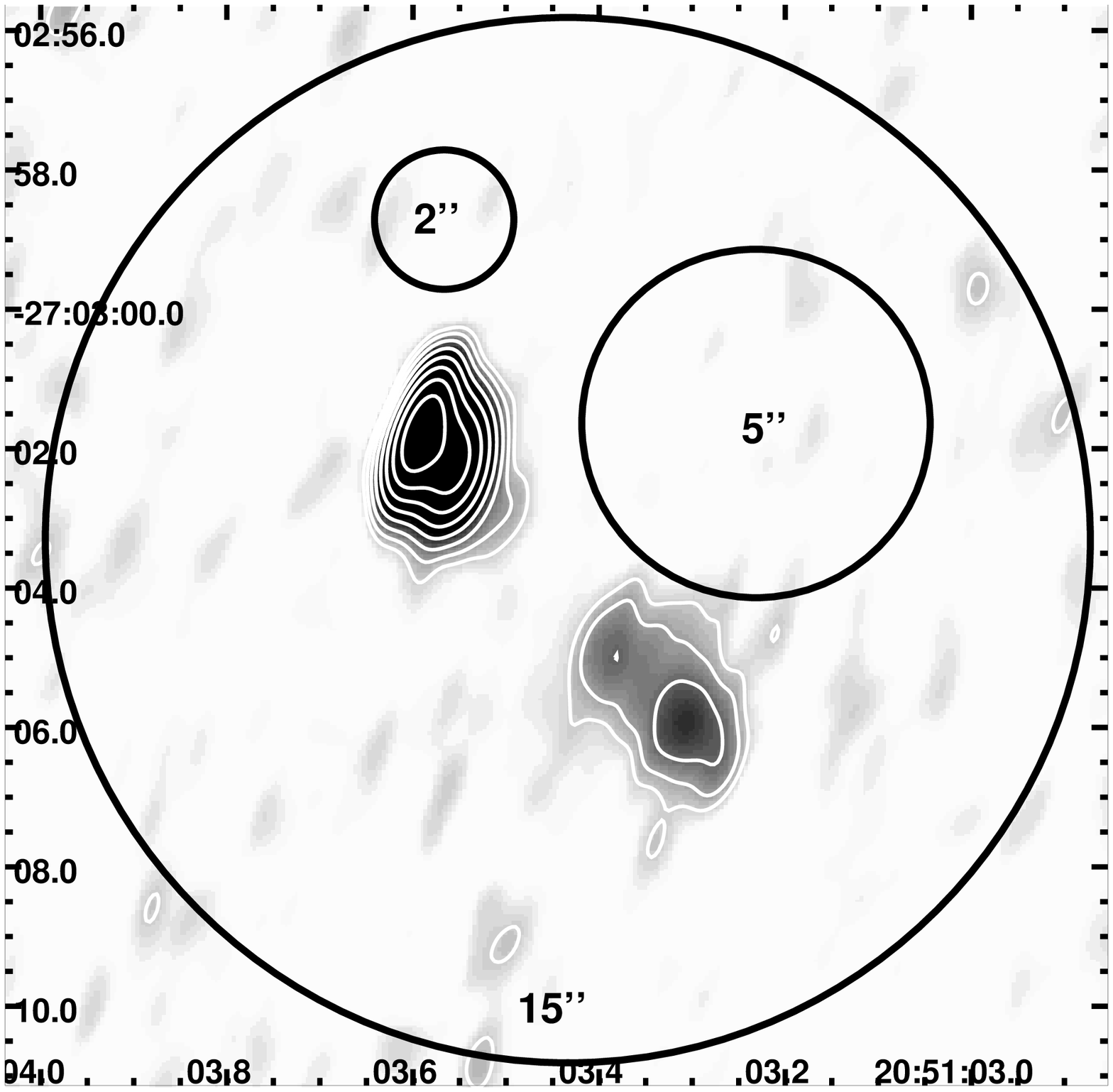}{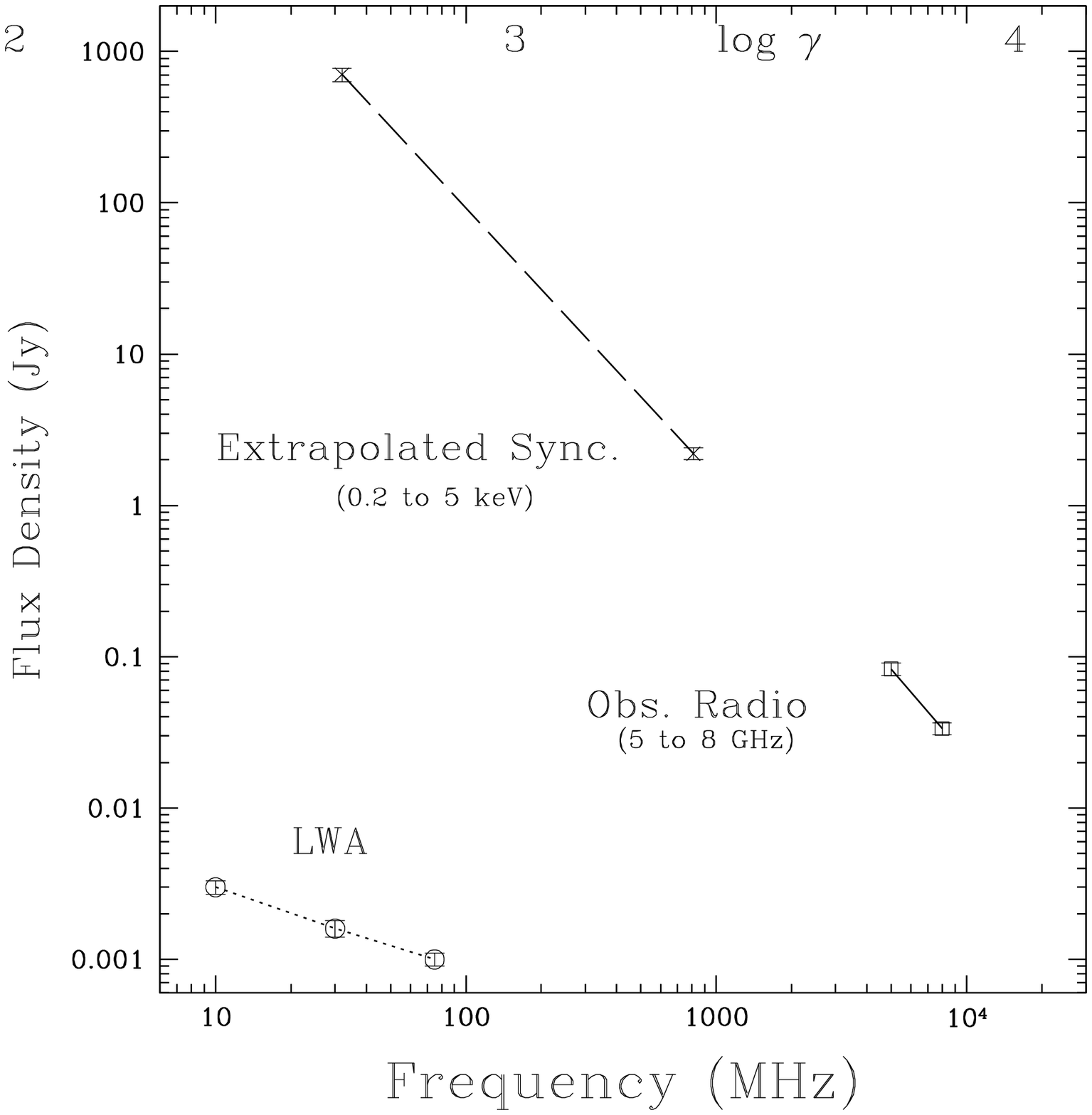}
\caption{The radio galaxy 2048-272 at z=2.06 (Overzier et
al. 2004).  The left panel shows a 5~GHz map from the VLA, again with
the predicted beamsizes from the LWA.  Contour levels start at 0.2mJy/beam
and increase by factors of two.  The right panel shows the spectrum, 
as in fig.~\ref{fig:351}.\label{fig:2048}}

\end{figure}

Another example of IC/CMB emission comes from the northern lobe of the
radio galaxy, 2048-272 at z=2.06.  This source is shown in
fig.~\ref{fig:2048} and it can be seen that at 10 MHz, the LWA beam
will include the southern lobe as well as the northern one.  This may
not preclude obtaining a spectrum for the northern lobe unless the
ratio of intensities is a strong function of frequency; a situation
that would be obvious from data above 30 MHz where the LWA can
separate the components.  An X-ray detection of the northern lobe, consistent
with IC/CMB emission, is reported in Overzier et al. (2004): 6$\pm$4
net counts.

Although the extrapolated spectrum ($\alpha_r$=1.79) is again centered
around $\gamma$=1000, because of the stronger B field, LWA will now be
sampling N($\gamma$) towards the bottom end of the segment responsible
for the 0.2 to 5 keV X-rays.  The LWA sensitivity provides a
comfortable margin to accommodate any turnover in the spectrum.  Thus
the IC/CMB model can be tested at z=2 where the energy density in the
CMB is 80 times larger than locally.

In addition to radio lobes, IC/CMB emission has been suggested from
cluster halo sources like Coma.  In the soft X-ray band we would again
be dealing with electrons with $\gamma\approx$1000, but the thermal
emission swamps the IC signal.  IC/CMB has also been suggested for EUV
excesses from clusters ($\gamma\approx$300) and for hard X-ray
excesses in the 20-80 keV band ($\gamma\geq~5000$, values similar to
the observed radio band).

\subsection{IC/CMB Emission from Jet Knots with Bulk Relativistic Velocities}

Although we take it as self evident that kpc scale jets have a bulk
relativistic velocity (with Lorentz factor
$\Gamma=(1-\beta^2)^{-\frac{1}{2}}$; $\beta=\frac{v}{c}$), from the
general one-sidedness of many radio, and essentially all detected
X-ray jets, there is no direct evidence supporting the large values of
$\Gamma$ (i.e. 10 and larger) required for the beaming model of
Tavecchio et al. (2000) and Celotti et al.  (2001).  Essentially all
of the models in the literature assume that N($\gamma$) can be
extrapolated down to the range from $\gamma$=10 to 300 with the power
law index, p=2$\alpha_r$+1, which is obviously based on the radio
spectral index measured at much higher frequencies and hence based on
larger electron energies.  If there were many more electrons than
predicted by the extrapolation, then smaller values of $\Gamma$ would
be indicated, and if there is a low energy cutoff of N($\gamma)$, a
much larger $\Gamma$ would be required, and consequently a smaller
viewing angle etc.  A true exponential cutoff would, of course make
the beaming IC/CMB model untenable since there would be no electrons
with the proper energy to produce the observed X-rays.

\begin{figure}[!ht]
\plottwo{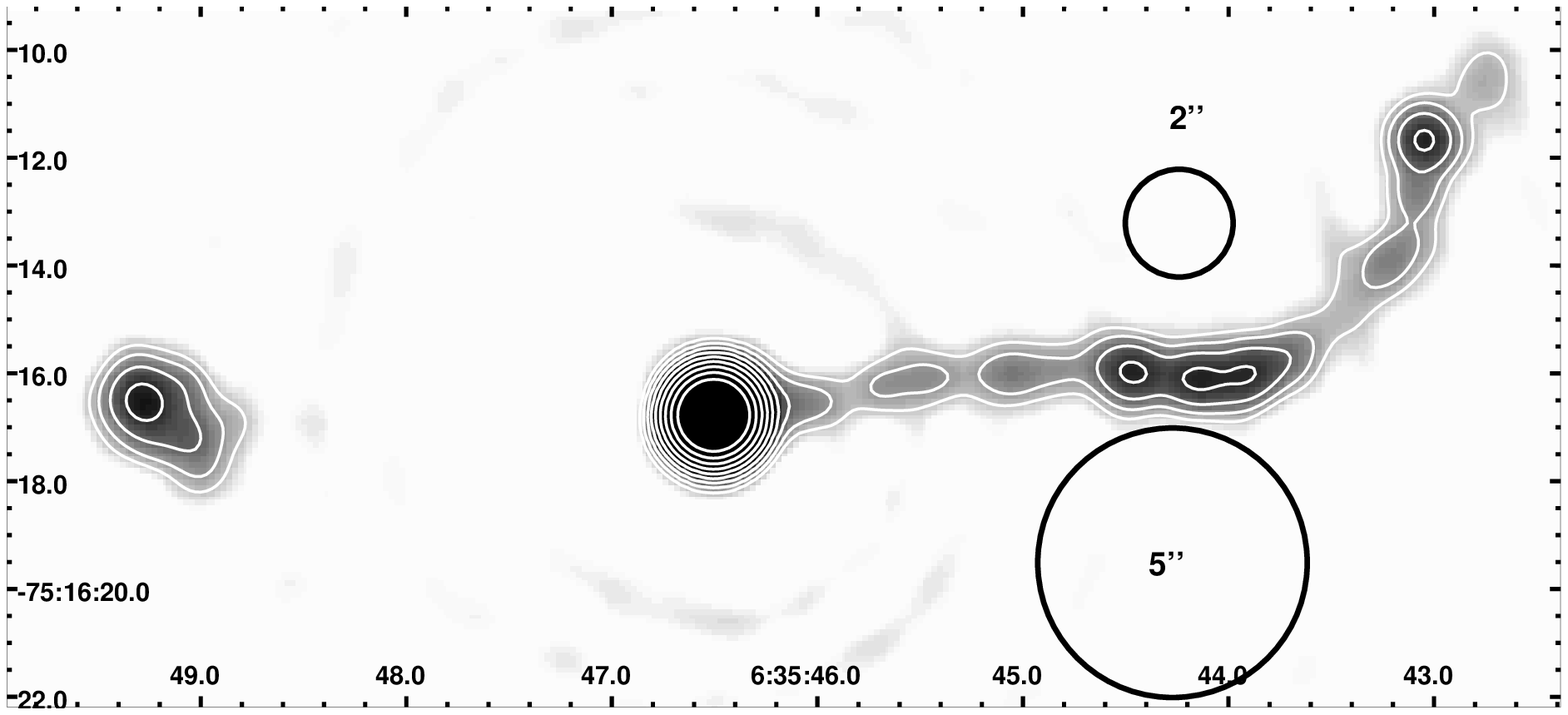}{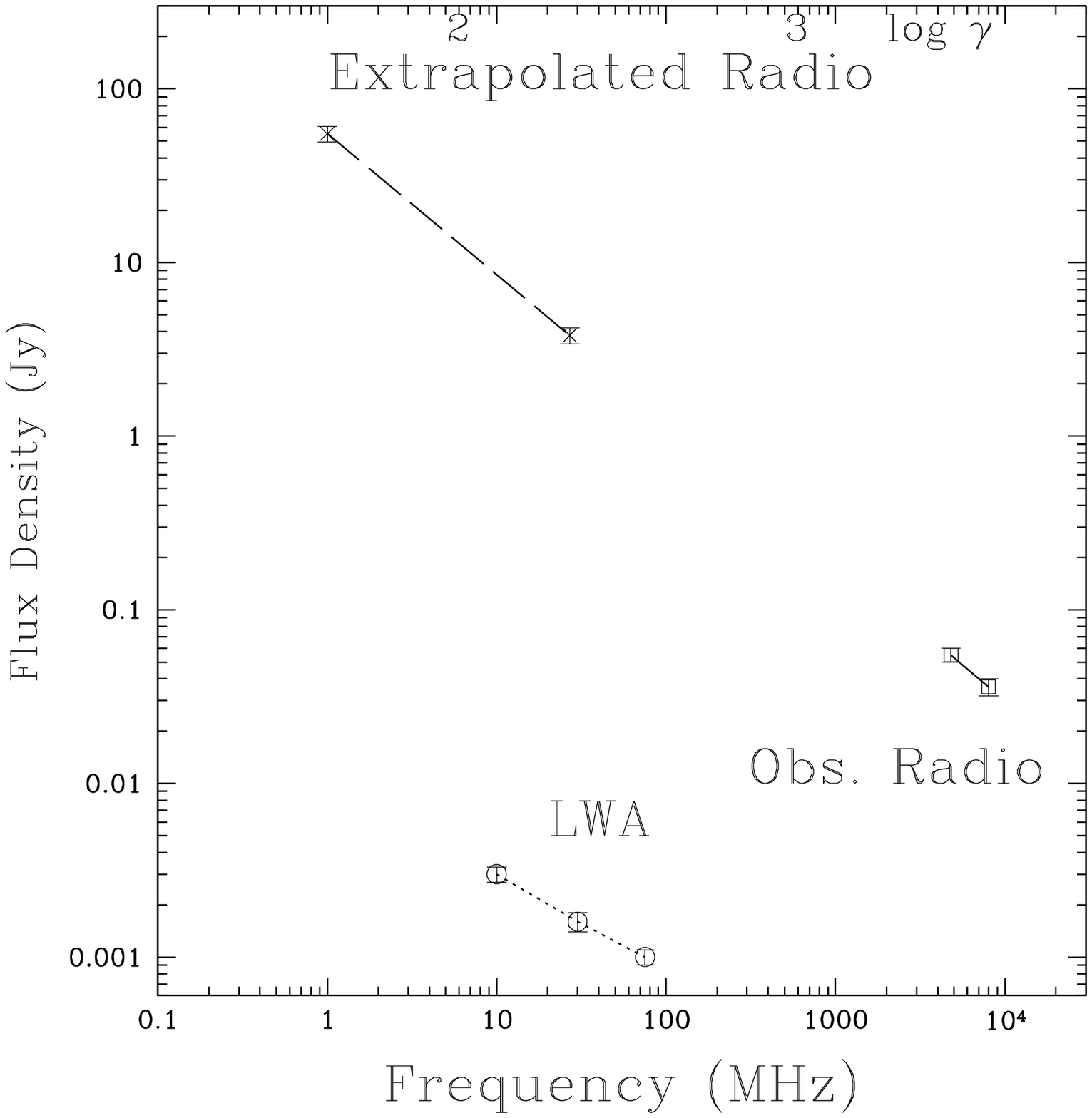}
\caption{PKS0637-752, a quasar at z=0.651.  The left panel shows a
5~GHz map from the ATCA (kindly provided by Lovell) with contours from
4 to 1024 mJy/beam increasing by factors of two.  The LWA beam widths
at 30 and 75 MHz are shown. The right panel shows the spectrum as for
previous figures.\label{fig:0637}}

\end{figure}

As an example, we show in fig.~\ref{fig:0637}, PKS0637-752, a quasar
at z=0.651.  Although not visible from New Mexico, it is perhaps, the
most widely believed example of IC/CMB X-ray emission from a jet
purported to have $\Gamma$=10.  Unless the core emission is absorbed
at low frequencies, the LWA resolution would limit a separation of
the jet knot from the core to frequencies of 30 MHz and greater.
Although the LWA would not sample electrons with $\gamma\approx$100
which provide the bulk of the observed X-rays, the 30 MHz data will
successfully bridge the gap in the extrapolation.  Thus the LWA data 
would provide a critical and convincing test of the IC/CMB beaming model
for X-ray emission from high power, kpc scale jets.

\subsection{IC Emission from Radio Lobes Illuminated by Quasar IR/Optical}

Brunetti (e.g. Brunetti et al. 2001) has suggested that if powerful
radio galaxies contain a hidden quasar as in the unified scheme, then
it may often happen that the cone of un-obscured quasar emission would
illuminate the radio lobes and for those parts of the lobes close to
the host galaxy, the IR/optical quasar emission would dominate the
local photon energy density.  This photon bath
would be anisotropic and hence there would be a preference for
scattering events to occur when an electron is moving towards the
core.  This leads to a gentle beaming of the IC X-rays so that the
receding side of the source should be brighter in the X-rays.  Since
the illuminating photons have frequencies of order 10$^{14}$ Hz, to
produce the observed X-rays, we need electrons with
$\gamma\approx$100.  Only a few examples of this behavior have been
given in the literature and actual numerical estimates depend on
assumptions about the unseen quasar flux.  Thus we will not pursue it
further except to state the obvious that once again, we have a
substantial extrapolation in $\gamma$ so that the LWA would provide
useful data so long as the angular resolution was sufficient.

\subsection{IC Emission from Radio Jets Illuminated by Hotspot Radio Emission}

Another rather specialized process for high luminosity radio galaxies
and quasars was suggested by Georganopoulos \& Kazanas (2003). For a
model in which a jet maintains a bulk relativistic velocity until it
decelerates at the terminal hotspot, the jet will experience a photon
energy density from the hotspot radiation amplified by $\Gamma^2$,
(the square of the jet's bulk Lorentz factor before the final
deceleration).

As the jet approaches the hotspot this photon energy density will
increase to a point where it will dominate other photon fields and be
sufficient to increase the ratio of IC to synchrotron emission.  The
resulting IC X-ray emission will be beamed along the jet direction,
and the electrons producing the X-ray emission will have $\gamma$'s
typical of those producing the normal band radio emission, so that no
extrapolations are required, and hence this process is not of direct
interest here.


\section{Conclusions}

\begin{itemize}

\item{Will LWA data provide constraints on acceleration mechanisms?}

	Yes - to the extent that we can determine $\gamma_{min}$

\item{Will these data test any current IC emission models?}

Yes - they will allow us to confirm or to disallow IC/CMB with
beaming.  If we can demonstrate a low E cutoff, this model will become
untenable .  The standard interpretation of IC/CMB X-rays from radio
lobes will also be tested.

\end{itemize}

The focus of this paper has been on the uncertainty involved in
extrapolating electron spectra to low energies, and the
ramifications of discovering that our assumed extrapolations may be
wrong.  A different outcome however, would be that LWA observations
convince us that instead, the extrapolations are valid and that there
is little or no change in the observed power law down to low values of
$\gamma$.  In that case, the IC X-ray data will provide measurements
of the amplitude of N($\gamma$) for lobes and clusters of galaxies,
and constraints on the amplitude and $\Gamma$ combination for jets.

For fairly clean examples of IC/CMB from lobes like 2038-272 and
3C351, we see that data from the LWA will permit us to sample the
electron spectra at the same energies that we observe with Chandra via
IC X-rays.  The IC data provide the amplitude, k$_e$, of N($\gamma$)
for various values of $\gamma$ below 1500.  The synchrotron degeneracy
between B and $k_e$ will thus be resolved and we will then have a good
estimate of $<B>$ {\itshape without the uncertainty of any
extrapolation}.  Furthermore, our estimates of the total energy (in
fields and particles), the non-thermal pressure, and the energy
density will be much more robust and thus we should begin to answer
the question of what is the value of $\gamma_{min}$.  The
equipartition constraint will have one less unknown, and pressure
balance arguments will then give us constraints on the filling factor
and the (1+k) factor (contribution from relativistic protons).

\acknowledgments

{We thank J. Eilek for discussions on the $\gamma_{min}$ problem.
This work was supported by NASA contract NAS8-39073, and grants
GO2-3144X, GO3-4142A, and GO3-4132X.}



\begin{thebibliography}{}



\bibitem[Braude et al. (1969)]{braude69}Braude, S. Ya., Lebedeva, O. M.,
Megn, A. V., Ryabov, B. P., and Zhouck, I. N. 1969, \mnras 143, 289

\bibitem[Brunetti et al. (2001)]{bru01}Brunetti, G., Cappi, M.,
Setti, G., Feretti, L., Harris, D.E. 2001, A\&A 372, 755

\bibitem[Carilli et al. (1991)]{cc91}Carilli, C. L., Perley, R. A.,
Dreher, J. W., \& Leahy, J. P. 1991, ApJ 383, 554

\bibitem[Celotti, Ghisellini, \& Chiaberge (2001)]{anna01}Celotti, A.,
Ghisellini, G., \& Chiaberge, M. 2001, MNRAS 321, L1-5

\bibitem[Eilek \& Hughes (1991)]{jean}Eilek, J. A. \& Hughes,
  P. A. 1991, in Beams and Jets in Astrophysics ed. \ Hughes,
  (Cambridge Astrophysics Series, No. 19, Cambridge University Press)
  p. \ 428

\bibitem[Georganopoulos \& Kazanas (2003)]{markos}Georganopoulos,
  M. \& Kazanas, D. 2003, ApJ 589, L5


\bibitem[Harris et al. (1988)]{4ssrs}Harris, D.E., Dewdney, P.E.,
Costain, C.H., McHardy, I., \& Willis, A.G. 1988, ApJ 325, 610

\bibitem[Harris, Carilli, \& Perley (1994)]{cyga}Harris, D.E.,
  Carilli, C.L. \& Perley, R.A. 1994, Nature 367, 713



\bibitem[Hardcastle et al. (2002)]{hotspots}Hardcastle, M. J.,
Birkinshaw, M., Cameron, R. A., Harris, D. E., Looney, L. W., \&
Worrall, D. M. 2002, ApJ 581, 948

\bibitem[Hardcastle et al. (2004)]{spots2}Hardcastle, M. J., Harris,
D. E., Worrall, D. M., \& Birkinshaw, M. 2004, ApJ 612, 729


\bibitem[Lesch \& Birk (1997)]{lb}Lesch, H. \& Birk, G. T. 1997, A\&A
324, 461

\bibitem[Overzier et al. (2005)]{5rg}Overzier, R. A., Harris, D. E.,
Carilli, C. L., Pentericci, L., Rottgering, H. J. A., \& Miley,
G. K. 2004, A\&A (submitted)

\bibitem[Pacholczyk (1970)]{andre}Pacholczyk, A.G. 1970, ``Radio
  Astrophysics'' W. H. Freeeman , San Franciso.


\bibitem[Roger, Bridle, \& Costain (1973)]{rbc}Roger, R. S., Bridle,
A. H., \& Costain, C. H. 1973, \aj 78, 1030

\bibitem[Tavecchio et al. (2000)]{tav}Tavecchio, F., Maraschi, L.,
Sambruna, R.M., \& Urry, C.M. 2000, ApJL 544, L23-26

\end{thebibliography}
\end{document}